\documentclass[%showpacs,
preprintnumbers,amsmath,amssymb]{revtex4}
\usepackage{graphicx}
\usepackage{dcolumn}% Align table columns on decimal point
\usepackage{bm}% bold math

\usepackage{graphicx}
\usepackage{epsfig}

\begin{document} %.........................................................
 \title{Variable viscosity condition in the modeling of a slider bearing}

 \author{Kedar Nath Uprety}
\email{kedar021@hotmail.com}

\affiliation{Central Department of Mathematics,\\Tribhuvan University,\\Kirtipur Campus, Kathmandu, Nepal}
%\pacs{02.30.Ik, 02.30.Hq, 03.65.Fd}
\author{Stefan C. Mancas}
\email{mancass@erau.edu}
\affiliation{Department of Mathematics, Embry-Riddle Aeronautical University,\\ Daytona-Beach, FL. 32114-3900, U.S.A}

%%%%%%%%%%%%%%%%%%%%%%%%%%%%%%%  AAA  %%%%%%%%%%%%%%%%%%%%%%%%%%
\begin{abstract}
\noindent %
To reduce tear and wear of machinery lubrication is essential. Lubricants form a layer
between two surfaces preventing direct contact and reduce friction between moving parts and
hence reduce wear. In this short letter the
lubrication of two slider bearings with parallel and nonparallel  is studied. First, we show that bearings with parallel plates cannot support any load. For bearings with nonparallel plates we are interested on how constant and temperature dependent viscosity affects the properties of the bearings.  Also, a critical temperature for which the bearings would fail due to excess in temperature is found for both latter cases. If the viscosity is constant, the critical temperature is given by an explicit formula, while for the non-constant viscosity the critical temperature can be always found from a closed form formula involving Weber functions.\\


\end{abstract}
%%%%%%%%%%%%%%%%%%%%%%%%%%%

\maketitle
%1111111111111111111111111111111111111111
\section{Slider bearing with parallel plates}\label{se1}
\begin{figure}  [!ht]%[x]%[!h]
   \centering
    \resizebox*{0.5\textheight}{!}{\includegraphics[]{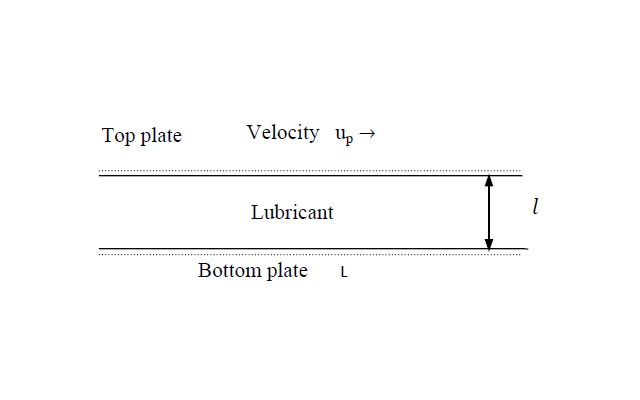}}
  \caption{Bearing with parallel plates}
  \label{Fig1}
 \end{figure}

The incompressible Navier-Stokes equations 
\begin{equation}\label{E1}
\frac{\partial \vec v}{\partial t} + \vec {v}\cdot \nabla \vec v + \frac{ \nabla p}{\rho}=\nu \Delta \vec v +\vec f
\end{equation}
together with the continuity equation
\begin{equation}\label{E2}
\frac{\partial \rho}{\partial t}+\nabla \cdot (\rho \vec v) = 0
\end{equation}
describe the motion of a fluid with velocity field $\vec v = (u, w)$, pressure  $p$, density  $\rho$,  kinematic viscosity $\nu$, and body forces $\vec f$ that may arise from friction \cite{Doe}. % is usually absent and other parameters remain constant.
Let the horizontal $x$ and vertical $z$ velocity components of the fluid be $u(x, z, t)$ and $w(x, z, t)$, and take incompressible fluid $\rho=const$ which gives $\nabla \cdot \vec v=0$,  respectively. Then from  (\ref{E1}) and (\ref{E2})  with $\vec f=\vec 0$, we obtain
\begin{align} \label{E3}
& \frac{\partial u}{\partial t} + u \frac{\partial u}{\partial x} + w \frac{\partial u}{\partial z} + \frac{1}{\rho} \frac{\partial p}{\partial x}
 = \nu \left(\frac{\partial^2 u}{\partial x^2} + \frac{\partial^2 u}{\partial z^2}   \right) \notag\\
 & \frac{\partial w}{\partial t} + u \frac{\partial w}{\partial x} + w \frac{\partial w}{\partial z} + \frac{1}{\rho} \frac{\partial p}{\partial z}
 = \nu \left(\frac{\partial^2 w}{\partial x^2} + \frac{\partial^2 w}{\partial z^2}   \right)
\end{align}  
\begin{equation} \label{E5}
  \frac{\partial u}{\partial x} + \frac{\partial w}{\partial z} = 0.
\end{equation}    
 We impose the following boundary conditions
  \[(u, w) = (0, 0)\,\, \mbox{at}\,\, z = 0 \,\, \mbox {and} \,\, (u, w) = (u_p, 0) \,\,\mbox{at}\,\,z = l \]
 where $u_p$ is the  horizontal velocity of the top plate, $l$ the separation distance between the plates along the $z$-axis, $L$ is the length of the plates, and the bottom plate is fixed, see Fig.\ref{Fig1}.
 
 %
%%%   FIGURE 1
%% %% =====================================================================
%%% 
Let a typical set of parameters for the bearing plate be $L= 5 \cdot \, 10^{-2} \, m$, $u_p= 1 \, m/sec$, $\rho=10^3 \,  kg/m^3$, $\mu=10^{-4} \, m^2/sec$. To non- dimensionalize  equations \eqref {E3}-\eqref{E5} we use new scaled  parameters  \[x = \overline{x} L,\,z = \overline{z} l ,\, u = \overline{u} u_p,\, w = \overline{w} \epsilon u_p,\, t = \overline{t} L/u_p, \, p=\bar p P, \] where $P$ is the undecided scaling  factor for pressure and will be determined later. By eliminating the terms which  have small
coefficients as  compared to  $1/\epsilon$, from (\ref{E3}) and (\ref{E5})
 we choose $P$ such that
 \begin{equation}\label{E6}
   P = \frac{ \mu  u_p L}{l^2},
 \end{equation} with $\mu=\rho \nu.$
 
Then  using \eqref{E6} we obtain
 \begin{equation}\label{E7}
  \frac{\partial \overline{p}}{\partial \overline{x}} = \frac{\partial^2 \overline{u}}{\partial \overline{z}^2} 
 \end{equation}  
  \begin{equation}\label{E8}
  \frac{\partial \overline{p}}{\partial \overline{z}} = 0 
 \end{equation}  
and from (\ref{E5}) we also have
 \begin{equation}\label{E9}
  \frac{\partial \overline{u}}{\partial \overline{x}} + \frac{\partial \overline{w}}{\partial \overline{z}} = 0 .
 \end{equation}   
 The new boundary conditions for $\overline{u}$ and $\overline{w}$ are
 \[\overline{u} = 0 \,\,\mbox{at}\,\, \overline{z} = 0 \,\,\mbox{and}\,\, \overline{u} = 1 \,\,\mbox{at}\,\, \overline{z} = 1\]
 \[\overline{w} = 0 \,\,\mbox{at}\,\, \overline{z} = 0 \,\,\mbox{and}\,\, \overline{w} = 0 \,\,\mbox{at}\,\, \overline{z} = 1\]
  One integration of (\ref{E8}) gives
  \[\overline{p} = \phi(\overline{x})\]
 and from (\ref{E7}) we obtain $\overline u_{\zeta}$ and $\overline u$ by two successive integrations
 \begin{equation}\label{E9a}
\frac{\partial \overline{u}}{\partial \overline{z}} = \frac{\partial \phi}{\partial \overline{x}}\overline{z} + c_1,
\end{equation}
and 
\begin{equation}\label{E10} 
 \overline{u} = \frac{\partial \phi}{\partial \overline{x}}\frac{\overline{z}^2}{2} + c_1\overline{z} + c_2.
 \end{equation}
 $c_1$ and $c_2$ are constants that depend on the boundary conditions $\overline u=0 \,\, \mbox{at}\,\, \overline z=0 $ and $\overline u=1 \,\, \mbox{at}\,\, \overline z=1 $ which give
 $c_2 = 0$ and 
 $ c_1 = 1 - \frac{1}{2}\frac{d\phi}{d\overline{x}}.$ Using all of these  the velocity in the $x$ direction is
\begin{equation}\label{E10a}
\overline{u} = \left(\frac{\overline{z}^2 - \overline{z}}{2}\right) \frac{d\phi}{d\overline{x}} + \overline{z}.
\end{equation}

 In the $z$ direction $\overline w$ is obtained easily from \eqref{E9} which gives 
  \begin{equation}\label{E9b}
   \frac{\partial \overline{w}}{\partial \overline{z}} = -\frac{\partial \overline{u}}{\partial \overline{x}}.
 \end{equation} 
 Using \eqref{E10a} in the above, by one integration we get 

   \begin{equation}\label{E9aa}
\overline{w}=\left(\frac{\overline z ^2}{4}-\frac{\overline z ^3}{6}\right)\frac{d^2 \phi}{d \overline x ^2}+c_3.
 \end{equation} 
As before, using the boundary conditions  $\overline w=0 \,\, \mbox{at}\,\, \overline z=0 \, \, \mbox{and} \,\, \overline z=1 $, we have $c_3=0$, and $c_3=-\frac{1}{12}\frac{d^2 \phi}{d \overline x ^2}$, which gives $\phi(\bar x)=c_4 \overline x+c_5$. But since $\phi=0$ then $\overline p=0$ and the velocity field is $\vec{v}=(z,0)$. Since the pressure is zero, the bearing with parallel plates cannot  support any load, see \cite{Pat, Up}. Therefore, we will consider next the case of nonparallel plates.

 \section{Slider bearing with nonparallel plates}

\begin{figure}  [!ht]%[x]%[!h]
   \centering
    \resizebox*{0.5\textheight}{!}{\includegraphics[]{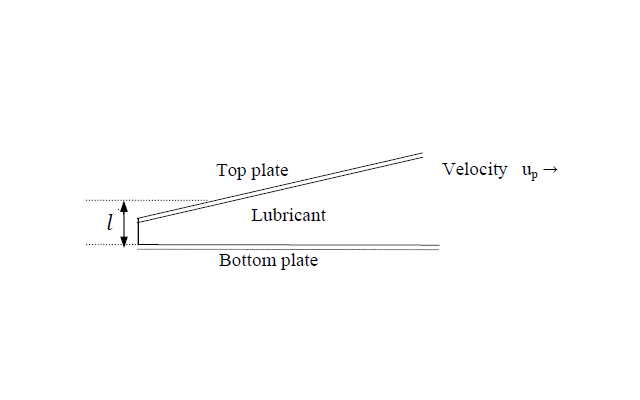}}
  \caption{Bearing with nonparallel plates}
  \label{Fig2}
 \end{figure}

 For the nonparallel plates, we have that bottom plate is flat and we assume that the top plate is a linear  function $z = h(x)$, see Fig. \ref{Fig2}.
 
 %
%%%   FIGURE 2
%% %% =====================================================================
 
Scaling the equations (\ref{E3})-(\ref{E5}) as before together with
\[ h(x)= l\overline{h}(x)\]
gives the same equations  (\ref{E7})- (\ref{E9}) but subject to new  boundary conditions
 \[\overline{u} = 0 \,\,\mbox{at}\,\, \overline{z} = 0 \,\, \mbox{and}  \,\,  \overline{u} = 1\,\, \mbox{at}\,\, \overline{z} = \overline{h}(\overline{x})\]

 From (\ref{E10})
 $c_2 = 0$ and 
 \[c_ 1= \frac{1}{\overline{h}(\overline x)} - \frac{1}{2} \frac{d \phi}{d \overline{x}} \overline{h}(\overline x).\]
 Thus 
\begin{equation}\label{E10aaa}
\overline u=\frac{d \phi }{d \overline{x}} \frac{\overline{z}^2-\bar{z}\bar{h}}{2} + 
  \frac{\overline{z}}{\overline{h}}
  \end{equation}
  From (\ref{E9}) then
 \begin{equation}\label{n}
  \frac{\partial \overline{w}}{\partial \overline{z}} = -\frac{\overline{z}^2 - \bar{z}\bar{h}}{2}\phi''(\bar{x})
   + \frac{\bar{z}}{2} \frac{d \bar{h}}{d\bar{x}} \phi'(\bar{x}) + \frac{\bar{z}}{\bar{h}^2}\frac{d \bar{h}}{d\bar{x}}
   \end{equation}
Integrating we obtain the velocity in the $z$ direction
\begin{equation}\label{E11}
 \bar{w} = -\left(\frac{\bar{z}^3}{6} - \frac{\bar{z}^2\bar{h}}{4}\right)\phi''(\bar{x}) + \frac{\bar{z}^2}{4} \frac{d\bar{h}}{d\bar{x}} \phi'(\bar{x}) + \frac{\bar{z}^2}{2\bar{h}^2}\frac{d \bar{h}}{d\bar{x}} + c_3
\end{equation}
By imposing the boundary conditions
\[\bar{w} = 0 \,\,\mbox{at}\,\, \bar{z} = 0\]
\[\bar{w} = 0 \,\,\mbox{at}\,\, \bar{z} = \bar{h}(\bar{x})\]
we have $c_3 = 0 $ and
\begin{equation}\label{E11b}
\frac{\bar{h}^3(\bar{x})}{12} \phi''(\bar{x}) + \frac{\bar{h}^2(\bar{x})}{4} \frac{d\bar{h}(\bar{x})}{d\bar{x}} \phi'(\bar{x}) 
   + \frac{1}{2}  \frac{d\bar{h}(\bar{x})}{d \bar{x}} = 0
   \end{equation}
which implies
\begin{equation}\label{E11bb}
\frac{d}{d\bar{x}}\left[ \frac{\bar{h}^3(\bar{x})}{12}\phi'(x) \right] + \frac{1}{2} \frac{d\bar{h}}{d\bar{x}} = 0.
   \end{equation}
 Assuming a linear profile \cite{Pat, Up} $\bar{h}(\bar{x}) = k_1\bar{x} + k_2$, and after simplification
 \begin{equation}\label{E12}
 \frac{(k_1\bar{x} + k_2)^3}{12} \phi'(\bar{x}) + \frac{1}{2} (k_1\bar{x} + k_2) = r_2
    \end{equation}
 then 
 \begin{equation}\label{E13}
 \phi'(\bar{x}) = \frac{-6}{(k_1\bar{x} + k_2)^2} + \frac{r_1}{(k_1\bar{x} + k_2)^3}.
  \end{equation}
One integration gives
 \begin{equation}\label{E14}
 \phi(\bar{x}) = \frac{6k_1}{k_1\bar{x} + k_2} + \frac{k_1r_1}{(k_1\bar{x} + k_2)^2} + r_2
  \end{equation}
  
\begin{figure}  [!ht]%[x]%[!h]
   \centering
    \resizebox*{0.25\textheight}{!}{\includegraphics[]{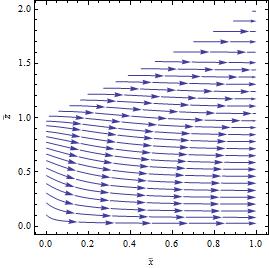}}
  \caption{Velocity field $\vec{v}$}
  \label{Fig3}
 \end{figure}

 The constants $r_1$ and $r_2$ are obtained by using boundary conditions  $\phi = 0$ at $\bar{x} = 0$ and $\bar{x} =1$. Hence,
 \begin{eqnarray}\label{r12}
 \left\{ \begin{array}{ll}
r_1 + r_2 = -6 \\
r_1 + 4r_2 = -12
\end{array} \right.
\end{eqnarray}
The linear system \eqref{r12} has solution $r_1 = -4, r_2 = -2$. Therefore the pressure 
 \[\bar{p} = \phi(\bar{x}) = \frac{2\bar{x}(1 - \bar{x})}{(1 + \bar{x})^2}\]
 and is positive for $\bar{x}\in(0, 1)$. Hence, the pressure is developed inside the fluid and the bearing supports a load given by   
 \begin{equation}\label{E15s}
  \mbox{load} = \int_{0}^{1} \frac{2\bar{x}(1 - \bar{x})}{(1 + \bar{x})^2} d\bar{x} = 6\ln(2) - 4
    \end{equation}
 Substituting the pressure  $\phi$ in (\ref{E10aaa}) and  (\ref{E11}) one can obtain an analytic expression for the velocity field $\vec{v}$ with components
 
   \begin{align}\label{E15n}
u&=\frac{\bar z}{(\bar x+1)^3}[\bar x(4\bar x-3\bar z+4)+\bar z] \notag \\
w&=\frac{2 \bar z^2}{(\bar x+1)^4}(\bar x-1)(\bar x-\bar z+1).
  \end{align}
 The velocity field $\vec{v}$ for the normalized scales, along with the streamlines are depicted in Fig.\ref{Fig3}.
  
  %%%   FIGURE 3
%% %% =====================================================================
 
  \subsection{Constant viscosity}
  In the following subsection we wish to find a critical temperature $T_c$ before the lubricant catches fire.   Suppose that the plates are kept at a constant temperature
  $T_1$ and lubricant catches fire at critical temperature $T_c$. We rescale the temperature according to 
\begin{equation}
T = T_1 + \bar{\theta} (T_c - T_1)
\end{equation}
 where  $\bar{\theta}$ is the non-dimensional variable $\in [0,1]$ and represents a critical parameter that will be determined. 

 The energy equation is
 \begin{eqnarray}\label{energ}
 &&\rho c_p \left( \frac{\partial T}{\partial t} + u\frac{\partial T}{\partial x} + w\frac{\partial T}{\partial z} \right)= \notag\\ 
   &&= \mu \left[ \left(\frac{\partial u}{\partial z} + \frac{\partial w}{\partial x}\right)^2 + 2 \left( \frac{\partial u}{\partial x}\right)^2
    + 2 \left( \frac{\partial w}{\partial z}\right)^2\right] + k\left( \frac{\partial ^2T}{\partial x^2} + \frac{\partial ^2T}{\partial z^2} \right)
 \end{eqnarray}   
  where $c_p$ is the specific heat of the viscous fluid and $k$ is its thermal conductivity. 
  
 The boundary conditions for $\bar{\theta}$ are
  \[\bar{\theta} = 0 \,\,\mbox{at}\,\, \bar{z} = 0 \,\,\mbox{and}\,\, \bar{z} = \bar{h}(\bar{x})\]   
As before, by neglecting the terms that are small as compared to $1/\epsilon$, it yields to the the new energy equation
   \begin{equation}\label{E15}
  \mu \frac{u_p^2}{l^2} \left(\frac{\partial \bar{u}}{\partial \bar{z}}\right)^2 
       = - \frac{k}{l^2}\frac{\partial ^2\bar{\theta}}{\partial \bar{z}^2}(T_c - T_1)
        \end{equation}
from which   
 \begin{equation}\label{E16}      
 \left(\frac{\partial \bar{u}}{\partial \bar{z}}\right)^2 = - B \frac{\partial ^2\bar{\theta}}{\partial \bar{z}^2}
 \end{equation}
 where
 \begin{equation}\label{bee}
 B = \frac{k}{\mu u_p^2} (T_c - T_1).
 \end{equation}
 Substituting $\bar{u}$, from (\ref{E10aaa})  then 
 \begin{equation}\label{E17}
\left[\frac{\partial }{\partial \bar{z}}\left( \frac{\overline{z}^2 - \overline{z}\overline{h}}{2} \frac{d\phi}{d\overline{x}} + \frac{\overline{z}}{\bar{h}}\right)\right]^2 = - B \frac{\partial ^2\bar{\theta}}{\partial \bar{z}^2}
 \end{equation}
 
 Integrating once and using boundary conditions for $\bar{\theta}$ we get 
\begin{equation}\label{E18}
-B\bar{\theta} = \phi'(\bar{x})^2 \frac{\bar{z}^4 -\bar{h}^2\bar{z}}{12} 
      + \left(\frac{1}{\bar{h}} - \frac{\bar{h}}{2}\phi'(\bar{x})\right)^2 \frac{\bar{z}^2 -\bar{h}\bar{z}}{2}
      + \phi'(\bar{x}) \left(\frac{1}{\bar{h}} - \frac{\bar{h}}{2}\phi'(\bar{x})\right) \frac{\bar{z}^3 -\bar{h}^2\bar{z}}{3} 
      \end{equation}
      
Hence,  for the linear profile  $h(\bar x)=\bar x+1$, and using \eqref{E13} the critical parameter becomes
\begin{equation}
\bar{\theta}=\frac{z}{3B(x+1)^6}[24x^2(x+1)^2z-(3x-1)^2z^3-8x(x+1)(3x-1)z^2+(x+1)^2(x^2-14x+1)]
\end{equation}
from which one can find the critical temperature $T_c$ of the lubricant before catches fire via \eqref{bee}.
 
\subsection{Variable viscosity}\label{s2}
Finally,  we will consider the lubricant in which  the viscosity decreases with the temperature. So (\ref{E3}) and (\ref{E5}) are
\begin{align}\label{E20}
&\frac{\partial u}{\partial t} + u \frac{\partial u}{\partial x} + w \frac{\partial u}{\partial z} + \frac{1}{\rho} \frac{\partial p}{\partial x}
 =  \frac{\partial }{\partial x}\left(\nu \frac{\partial u}{\partial x}\right)
     + \frac{\partial }{\partial z}\left(\nu \frac{\partial u}{\partial x}\right) \notag \\
 & \frac{\partial w}{\partial t} + u \frac{\partial w}{\partial x} + w \frac{\partial w}{\partial z} + \frac{1}{\rho} \frac{\partial p}{\partial z}
 =  \frac{\partial }{\partial x}\left(\nu \frac{\partial w}{\partial x}\right)
     + \frac{\partial }{\partial z}\left(\nu \frac{\partial w}{\partial x}\right)
\end{align}  

Using dimensionless viscosity $\nu = \nu_0\bar{\nu}$ in 
\begin{equation}\label{E22}
 \frac{\partial \bar{p}}{\partial \bar{z}} = \frac{\partial }{\partial \bar{z}}\left(\nu \frac{\partial \bar{u}}{\partial \bar{z}}\right) 
\end{equation}
we obtain
\begin{equation}
 \frac{\partial \bar{p}}{\partial \bar{z}} = \nu_0\frac{\partial }{\partial \bar{z}}\left(\bar{\nu} \frac{\partial \bar{u}}{\partial \bar{z}}\right),
\end{equation} 
and the energy equation reduces to
  \begin{equation} \label{E25}     
   \bar{\nu}\left(\frac{\partial \bar{u}}{\partial \bar{z}}\right)^2 = - B \frac{\partial ^2\bar{\theta}}{\partial \bar{z}^2}
 \end{equation}
 with B given by \eqref{bee}.
 
As the viscosity changes with temperature we will assume $\bar{\nu} = \alpha/\bar{\theta}$. From (\ref{E22}) as $p = \phi(\bar{x})$ we have
  \begin{equation} \label{E26}   
    \frac{\partial \bar{u}}{\partial \bar{z}} = \frac{\phi'(\bar{x})\bar{z} + c_1(\bar{x})}{\bar \nu}
     \end{equation}
or in terms of $\bar {\theta}$
    \begin{equation} \label{E27}   
\frac{\partial ^2\bar{\theta}}{\partial \bar{z}^2} = - \frac{\bar{\theta}}{\alpha B} \left( \phi'(\bar{x})\bar{z} + c_1(\bar{x}) \right)^2
   \end{equation}
We recognize \eqref{E27} as a parabolic differential equation which will be solved using Weber functions. To see that, let us the the transformation 
\begin{equation}\label{tr}
\zeta=\frac{\sqrt 2 \phi(\bar x)}{\sqrt[4]\alpha \beta}\left (\bar z +\frac{c_1(\bar x)}{\phi'(\bar{x})}\right)
\end{equation}
 which leads to the special case of Weber\rq{}s equation \cite{Steg}
\begin{equation}\label{wb}
\bar \theta_{\zeta \zeta}+\left( n+\frac 12 -\frac 1 4 \zeta^2\right)\bar \theta=0
\end{equation} provided that $n=-\frac 1 2$.
Since we have two independent solutions 
\begin{align}\label{s1}
\bar \theta_{1}&=e^{-\frac{\zeta^2}{4}} {}_1F_1\Big(\frac1 4, \frac 1 2;\frac {\zeta^2}{2}\Big) \notag \\
\bar \theta_{2}&=\zeta e^{-\frac{\zeta^2}{4}} {}_1F_1\Big(\frac3 4, \frac 3 2;\frac {\zeta^2}{2}\Big) 
\end{align}
where $ {}_1F_1(a,b;\zeta)=\sum_{n=0}^{\infty}\frac{(a)_n}{(b)_n}\frac{\zeta^n}{n!}$ is the confluent hypergeometric function that satisfies 
\begin{equation}\label{ku}
\zeta y\rq{}\rq{}+(b-\zeta)y\rq{}-a y=0
\end{equation}
one can construct the auxiliary functions \cite{Whi} from \eqref{ku}
\begin{align}\label{s2}
Y_{1}&= \frac {\sqrt \pi}{\Gamma(1/4)}\bar \theta_{1}\notag \\
Y_{2}&=\frac {2 \sqrt \pi}{\Gamma(1/4)}\bar \theta_{2}
\end{align} which in turn yield to the Whittaker functions

\begin{align}\label{s3}
U(\zeta)&=\frac{\sqrt 2}{2}(Y_1-Y_2)\notag \\
V(\zeta)&= \frac{1}{\Gamma(1/2)}\frac{\sqrt 2}{2}(Y_1+Y_2)
\end{align}
Using all of the above, the solution to \eqref{E27} may be written as
\begin{equation}
\bar \theta(\zeta)=\frac{\sqrt \pi e^{-\frac{\zeta^2}{4}}}{\sqrt[4]{2}\Gamma(3/4)}\left( {}_1F_1\Big(\frac1 4, \frac 1 2;\frac {\zeta^2}{2}\Big)-\sqrt 2 \zeta  {}_1F_1\Big(\frac3 4, \frac 3 2;\frac {\zeta^2}{2}\Big) \right)
\end{equation}
which upon using the substitution again to go back to the $\bar z$ variable using \eqref{tr}, together with the boundary conditions for $\bar{\theta}$ that determine $c_1(\bar x)$ would yield the critical temperature before the bearing will fail due to excess in heath.
\section{Conclusion}

In this short letter we studied two models of slider bearings with nonparallel plates and two types of viscous fluids between the  plates.  One of the plates is fixed and
other is moving horizontally with a constant velocity. The work obtained due to the motion of the top plate on  lubricant makes the temperature of the viscous layer change, which yields in the properties of the viscous fluid to be affected. We found the conditions under which the safe operation of the bearing is ensured. That is, we found two critical temperatures before the lubricant fails. The first temperature was based on the simpler case of constant viscosity, while the second case was found from the temperature dependent viscosity. Closed form solutions in terms of Weber functions were found for the non-constant case. These findings can be extended to slider bearings with different geometries $h(\bar x)$, and viscosities that depend on temperature differently from the discussion here. 
\section*{ACKNOWLEDGMENTS}
%If there are any acknowledgments to research support etc., they go here. Notice that this section is unnumbered. 
The first author would  like to acknowledge to D. P. Patel, I. Pendharkar, S. Chapman, A. Fitt with whom he worked with in the program:`` Regional International Workshop on Industrial Mathematics\rq{}\rq{}, IIT Bombay, 2-6 December 2006, organized by UNESCO.

\end{document}